\documentclass{aa}
\pdfoutput=1
\usepackage{graphicx, hyperref}
\usepackage{natbib,epsfig,pslatex,amsmath,txfonts}
\usepackage[normalem]{ulem}
\begin{document}
\title{Characterizing a cluster's dynamic state using a single epoch of radial velocities}
\author{Michiel Cottaar \inst{1}
	\and Michael R. Meyer \inst{1}
	\and Richard J. Parker \inst{1}}
\institute{Institute for Astronomy, ETH Zurich, Wolfgang-Pauli-Strasse 27, 8093 Zurich, Switzerland}
\date{\today}

\titlerunning{Characterizing a cluster's dynamic state}

\abstract{Radial velocity measurements can be used to constrain the dynamical state of a stellar cluster. However, for clusters with velocity dispersions smaller than a few km\,s$^{-1}$ the observed radial velocity distribution tends to be dominated by the orbital motions of binaries rather than the stellar motions through the potential well of the cluster.}
{Our goal is to characterize the intrinsic velocity distribution of a cluster from a single epoch of radial velocity data even for a cluster with a velocity dispersion of a fraction of a km\,s$^{-1}$.}
{We investigate a maximum likelihood procedure, which was pioneered separately by Odenkirchen et al. (2002) and Kleyna et al. (2002). Assuming a period, mass ratio, and eccentricity distribution for the binaries in the observed cluster this procedure fits a dynamical model describing the velocity distribution for the single stars and center of masses of the binaries, simultaneously with the radial velocities caused by binary orbital motions, using all the information available in the observed velocity distribution. We test the capability of this procedure to reproduce the velocity dispersion of an observed cluster, using radial velocity data of an open cluster and Monte Carlo simulations.}
{We find that the fits to the intrinsic velocity distribution depend only weakly on the binary properties assumed, so the uncertainty in the fitted parameters tends to be dominated by statistical uncertainties. Based on a large suite of Monte Carlo simulations we provide an estimate of how these statistical uncertainties vary with the velocity dispersion, binary fraction, and the number of observed stars, which can be used to estimate the sample size needed to reach a specific accuracy. Finally we test the method on the well-studied open cluster NGC 188, showing that it can successfully reproduce a velocity dispersion of only 0.5\,km\,s$^{-1}$ using a single epoch of the multi-epoch radial velocity data.}
{If the binary period, mass ratio, and eccentricity distribution of the observed stars are roughly known, this procedure can be used to correct for the effect of binary orbital motions on an observed velocity distribution. This allows for the study of the dynamical state of a stellar cluster with a small velocity dispersion from a single epoch of radial velocity data.}

	
\maketitle

\section{Introduction}
Most stars form in groupings or clusters, which are bound whilst still embedded in the birth molecular cloud and occasionally ($\sim 10$\%) even after the gas has been expelled \citep{Lada03}. In such an environment gravitational interactions between stars could dynamically process the binaries in the cluster, as well as disturb any disks or planetary systems orbiting the stars \citep[e.g.][]{Kroupa95,Adams06,Parker11a}. To fully understand the formation and evolution of these stellar clusters, as well as their effect on individual stars and multiple systems, we have to study their dynamical properties. 

For a resolved cluster with extensive photometric observations, we can measure the surface density of the cluster. Assuming a density distribution along the line of sight for the stars (e.g. assuming a spherical cluster) we can use this to calculate the potential energy of the cluster. Combined with the radial velocity dispersion measured for an unbiased sample of stars, the cluster's kinetic energy can be calculated. The ratio of the kinetic to potential energy can be used to determine whether a cluster is either expanding, is collapsing, or is in virial equilibrium and hence whether the cluster will be stable on the dynamical timescale. 

If the radial velocities of a large number of stars is available a more sophisticated dynamical model based on the potential energy and density profile of the cluster can be tested. In such a model the velocity distribution is not characterized by a single velocity dispersion, but by a velocity distribution which varies with mass and/or distance from the cluster centre and the velocity distribution could be non-Gaussian. In globular clusters, measurements of the radial dependence of the velocity dispersion have been used to for example show consistency of the stellar dynamics with the \citet{King65} thermal equilibrium model \citep{Gunn79}, to constrain the anisotropy of the cluster \citep[e.g.][]{van-de-Ven06, Sollima12} and even to check for deviations from Newtonian gravity \citep[e.g.][]{Baumgardt05,Sollima12}. These studies were aided by the relatively high velocity dispersions and the relatively low binary fractions \citep[e.g.][]{Gunn79,Davis08} of most globular clusters. 

These studies can not be simply extended to lower mass clusters because these clusters tend to have smaller velocity dispersions. For small velocity dispersion (i.e. less than a few km\,s$^{-1}$) the radial velocities from binary orbital motions can inflate the measured velocity dispersion of a cluster by many km\,s$^{-1}$ \citep[e.g.][]{Kouwenhoven08,Kouwenhoven09,Gieles10,McConnachie10}. The importance of these binary orbital motions depends on the original velocity dispersion, the binary fraction and the period, mass ratio, and eccentricity distributions of the binaries in the cluster. These velocities induced by binary orbital motions completely mask the very small velocity dispersions  ($< 1\,{\rm km\,s}^{-1}$) expected in open clusters \citep[e.g.][]{Geller08,Geller10} and possibly in ultra-faint dwarf spheroidals \citep[e.g.][]{McConnachie10}. Even the larger velocity dispersions of young massive clusters \citep[e.g.][]{Bosch09,Gieles10,Cottaar12,Henault-Brunet12a}, local star-forming regions \citep[e.g.][]{Tobin09}, and some low-mass globular clusters \citep[e.g.][]{Odenkirchen02,Blecha04,Sollima12} can be inflated by binaries. 

\begin{figure}
	\begin{center}
 		\includegraphics[width=.5\textwidth]{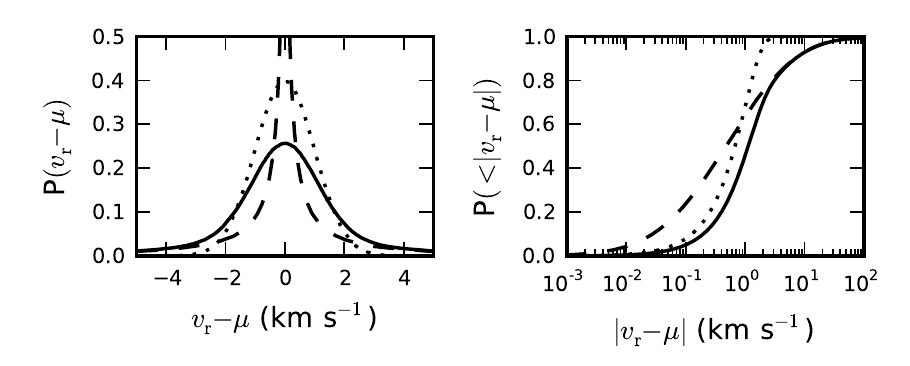}
		\caption{\label{fig:cdf} Probability distribution function (left) and cumulative distribution function (right) for radial velocity offsets from the mean velocity for velocities drawn from a Gaussian distribution with a dispersion of 1\,km\,s$^{-1}$ (dotted, eq \ref{eq:punder}), the velocity offsets expected due to binary orbital motions (dashed, eqs \ref{eq:vprim} and \ref{eq:pbin}), and the velocity distribution for a cluster with a velocity dispersion of 1\,km\,s$^{-1}$ and a 100\% binary fraction (solid), which is the convolution of the dotted and dashed distribution (eq \ref{eq:ptotal}). Note that the binary orbital motions have a significant chance of giving a velocity offset of only tens of m/s, leading to a very high central peak in the probability density function, which has been cut off in the left panel.}
	\end{center}
\end{figure}

In Figure \ref{fig:cdf} we have illustrated the effect of binary orbital motions on the probability density function and cumulative distribution of a single epoch of radial velocity data. The dotted line traces a Gaussian with a width of 1\,km\,s$^{-1}$, representing the velocity distribution of single stars and the center of masses of binaries in the observed cluster. The dashed line shows the probability distribution of radial velocity offsets induced by binary orbits under several assumptions about the binary properties discussed in Sec. \ref{sec:bin}. The solid line finally shows the convolution of these distributions, representing the distribution from which the observed radial velocities will be drawn, assuming a binary fraction of 100\%. The binary orbital motions both broaden the Gaussian distribution and add a high-velocity tail to the distribution. The goal in this paper is to present a procedure which can be used after measuring a single epoch of radial velocities (having a distribution similar to the solid line) to characterize the intrinsic velocity distribution (dotted), assuming the binary period, mass ratio, and eccentricity distributions, which corresponds to fixing the shape of the dashed line. This fitting can accurately reproduce the intrinsic velocity distribution, because the velocities added due to binary orbital motions will generally have a very different velocity distribution from the intrinsic velocity distribution.

An often used method to deal with the binary contamination is to obtain multi-epoch observations, which can be used to distinguish between single and binary stars \citep[e.g.][]{Geller08,Geller10,Tobin09,Cottaar12}. The single stars (in this context generally defined as stars for which no significant radial velocity variations were found) as well as any binaries for which the velocity of the center of mass has been measured, can then be used to determine the velocity distribution of the cluster. However even after a large number of observations, there might still be undetected binaries, which broaden the measured velocity distribution. For example a circular solar equal-mass binary with a period of a 1000 years, still has a velocity amplitude of 2\,km\,s$^{-1}$. 

In absence of multi-epoch data or to correct for any remaining undetected binaries Monte-Carlo simulations can be used to estimate for which velocity dispersion the broadening due to an assumed binary population would lead to the observed width of the distribution \citep[e.g.][]{Gieles10,McConnachie10,Sollima12}. These methods tend to focus on reproducing only the width of the distribution, rather than the full velocity distribution observed.

We investigate here an alternative strategy, which was used by \citet{Odenkirchen02} in their study of the globular cluster Palomar 5 and separately by \citet{Kleyna02} in their study of the Draco dwarf spheroidal. If the binary period, mass ratio, and eccentricity distributions are roughly known for the type of cluster being studied, this can be used to simultaneously fit the intrinsic velocity distribution and the fraction of binaries. The high-velocity tail caused by the orbital motions of close binaries can be used to determine the binary fraction in the observed cluster, with the distribution of velocities in the tail acting as a test on the assumed binary properties. From the high-velocity tail the effect of these binaries can then be extrapolated to lower velocities to quantify the degree with which the velocity distribution has been broadened by binaries.

In Sec. \ref{sec:method} we describe the maximum-likelihood analysis we use to fit the intrinsic velocity distribution and the binary orbital velocity distribution. In Sec. \ref{sec:accuracy} we discuss the accuracy of the method, computing the statistical uncertainty as a function of the velocity dispersion, binary fraction, and number of observed stars, as well as the systematic offsets that can be introduced by differences between the assumed and true binary properties. Here we will also provide a tool to estimate the sample size needed to measure the dynamical state or binary fraction of a cluster with a desired precision prior to the observations. We test our model on the old open cluster NGC 188 in Sec \ref{sec:ngc188}. \citet{Geller08,Geller09,Geller11} present multi-epoch radial velocity data for a large number of stars in this cluster and found that the single stars in their data set had a velocity dispersion of $0.49^{+0.07}_{-0.08}$\,km\,s$^{-1}$. We show that we can reproduce their measured velocity dispersion using only a single epoch of their data. We discuss the limitations of the method and a possible extension to correct for any undetected binaries in multi-epoch data in Sec. \ref{sec:disc} and summarize our results in Sec. \ref{sec:summary}.

\section{Method\label{sec:method}}
We consider a single epoch of $N_{\rm obs}$ radial velocity observations $v_{\rm obs,\ i}$ with measurement uncertainties of $\sigma_{\rm obs,\ i}$ of stars with mass estimates $m_i$. Our goal is to fit a dynamical model to the binary-corrected velocity distribution of the stars in the cluster for a set of assumptions about the binary period, mass ratio, and eccentricity distributions. As free parameters in this fit we will use the binary fraction in the cluster as well as any parameters used in the dynamical model (e.g. mean velocity, velocity dispersion). 

First we calculate the likelihood function, which gives the likelihood of reproducing the observations given the parameters. This likelihood function is unique for every star, as it depends on the measurement error, the mass estimate of the star, and possibly other stellar properties taken into account in the dynamical model (Sec. \ref{sec:vdist}). Then we use the maximum likelihood estimator to convert the likelihood to reproduce the observations as a function of the parameters into a probability distribution of the parameters given the observed radial velocities (Sec. \ref{sec:fit}).

\subsection{Incorporating binaries in the velocity distribution\label{sec:vdist}}
The observed velocities can be seen as randomly drawn from the dynamical model ($v_{\rm dyn}$) describing the intrinsic velocity distribution. We will take the measurement uncertainty into account directly in this dynamical model. So for the single stars this velocity will match the observed velocity. For a random subset of stars (the binaries) we add an additional velocity due to binary orbital motions ($v_{\rm bin}$). The probability for each star to be a binary depends on the binary fraction of the cluster and the distribution of the added velocities depends on the assumed period, mass ratio, and eccentricity distribution as well as the mass of the star. To calculate the probability of observing a certain velocity ($v_{\rm obs}$) for a star in a binary, we have to integrate over the probability of any combination of $v_{\rm dyn}$ and $v_{\rm bin}$, for which $v_{\rm dyn} =v_{\rm obs} - v_{\rm bin}$. This corresponds to taking the convolution of the distributions of $v_{\rm dyn}$ and $v_{\rm bin}$. The full velocity distribution is given by the sum of the distributions for single and binary stars:
\begin{equation}
\begin{split}
  \mathcal{L}_i&(v_{\rm obs,\ i}) = (1-f_{\rm bin}) \mathcal{L}_{\rm dyn,\ i}(v_{\rm obs,\ i}) + \\
  &f_{\rm bin} \int_{-\infty}^{+\infty}{\mathcal{L}_{\rm dyn,\ i}(v_{\rm obs,\ i}-v_{\rm bin}) \cdot \mathcal{L}_{\rm bin,\ i}(v_{\rm bin})}\, \mathrm{d} v_{\rm bin}, \label{eq:ptotal}
\end{split}
\end{equation}
which gives the likelihood of measuring a velocity $v_{\rm obs,\ i}$ as a function of the binary fraction ($f_{\rm bin}$) and any other free parameters in the dynamical model. The subscript $i$ is added to the likelihood function ($\mathcal{L}_i$) to emphasize that this likelihood function is different for every star due to its dependence on the measurement error and the mass of the star.  We note that in the derivation of equation \ref{eq:ptotal} we assumed that the single stars and the center of masses of the binaries have the same dynamical distribution. Mathematically this assumptions is unnecessary and for some dynamical models it might be useful to drop this assumption (e.g. dynamical models that include mass segregation, where the velocity distribution depends on the total mass of a binary). 

In equation \ref{eq:ptotal} we have not taken any triple or higher-order systems into account. For solar mass these higher-order systems are rare \citep{Raghavan10}. However, for more massive primaries the fraction of triple and higher-order system is higher \citep{Zinnecker07,Raghavan10}. For the system to be stable the radial velocity of any star will still be dominated by a single (generally the closest) companion, which could be either a single star or a multiple system with a much smaller semi-major axis. This implies that the radial velocity offsets expected for a star in triple and higher-order systems can to first order still be approximated by that of a binary.

In this paper we will focus on measuring the velocity dispersion of a cluster. Using the maximum likelihood estimator the standard deviation and mean of a set of observed velocities can be derived by maximizing the likelihood that the observations were drawn from a Gaussian distribution \citep{Pryor93}. So in order to measure the mean velocity and velocity dispersion corrected for orbital motions of the stars in a cluster, we will use a Gaussian model to fit the intrinsic velocity dispersion
\begin{equation}
  \mathcal{L}_{\rm dyn,\ i}(v_{\rm dyn})= \frac{1}{\sqrt{2 \pi (\sigma_i^2+\sigma_c^2)}} \exp\left(-\frac{(v_{\rm dyn}-\mu_c)^2}{2 (\sigma_i^2+\sigma_c^2)}\right), \label{eq:punder}
\end{equation}
which gives the likelihood for a velocity $v_{\rm dyn}$ given a measurement error $\sigma_i$ and the mean velocity $\mu_c$ and velocity dispersion $\sigma_c$ of the cluster. 

If the observed stars have not been drawn randomly from the cluster, the velocity dispersion of the observed stars is no longer representative for the velocity dispersion of the whole cluster. As most spectroscopic observations will focus on the brightest stars, there will generally be a deviation between the observed velocity dispersion of these stars and the velocity dispersion of the cluster, if the cluster is mass segregated. When the observed stars are not representative of the whole population, a more realistic model could still be fitted. In this case the velocity dispersion ($\sigma_c$) should no longer taken to be a constant for all stars, but the dependence of the velocity dispersion on the distance from the cluster center or the mass of the star should be modeled. Another possible extension would be to model rotation, by allowing $\mu_c$ to vary across the cluster. In our efforts to calculate the accuracy with which the intrinsic velocity distribution can be retrieved (Sec. \ref{sec:accuracy}) we will focus on the simple case with only a Gaussian intrinsic velocity distribution and hence on the accuracy with which we can measure a single velocity dispersion. In Sec. \ref{sec:ngc188} we will have to adjust the model for the intrinsic velocity distribution when fitting the radial velocity data from NGC 188 to take into account the contamination of fore- and background stars in the observed sample.

Given a set of assumptions about the binary period, mass ratio, and eccentricity distributions, we numerically calculate the distribution of radial velocity offsets due to binary orbital motions. We start by randomly drawing a large number ($\sim 10^5$) of binaries with period $P_j$, mass ratio $q_j$ and eccentricity $e_j$ from the assumed distributions. In addition to these we draw for every binary a random time $t_j$ since the last passing of periastron. To minimize the time of the computation, we have pre-computed a dense two-dimensional grid giving the relative velocity between the binary components for various eccentricities ($0 \le e_j < 1$) and phases ($0 \le t_j/P_j < 1$) for a period of 1 year and a semi-major axis of 1 AU. The velocity taken from this grid can be used to calculate the velocity of the observed star through
\begin{equation}
  v_j=\left(\frac{m_i}{{\rm M}_{\odot}}\right)^{\frac{1}{3}} \left(\frac{{\rm yr}}{P_j}\right)^{\frac{1}{3}} \frac{q_j}{(1+q_j)^{\frac{2}{3}}} v_{\rm grid}(e_j,t_j/P_j), \label{eq:vprim}
\end{equation}
where $m_i$ is the mass of the observed star, and $v_{\rm grid}$ is the velocity calculated for the closest match in eccentricity and phase from the table.

In order to calculate the distribution of velocity offsets, we still have to take into account the projection of the binary motions along the line of sight. Such a projection for a star with an absolute velocity offset $v_j$ in three dimensions will give a uniform probability between $-v_j$ and $+v_j$ to get a certain velocity offset along the line of sight. This flat distribution implies that in the calculation of the total probability to have a velocity offset due to binary motions of $v_{\rm bin}$, the contribution of all binaries with $v_j < |v_{\rm bin}|$ is zero, while all binaries with $v_j > |v_{\rm bin}|$ contribute a probability of $\frac{1}{2 v_j}$. For the list of primary velocities obtained (eq. \ref{eq:vprim}) this gives for the likelihood of binary-induced motions to produce a radial velocity offset $v_{\rm bin}$
\begin{equation}
  \mathcal{L}_{\rm bin ,\ i} (v_{\rm bin}) = \sum_{j; v_j > |v_{\rm bin}|} \frac{1}{2 v_j}, \label{eq:pbin}
\end{equation}
where we sum the $\frac{1}{2 v_j}$ for all $v_j>|v_{\rm bin}|$. The velocity distribution obtained using equations \ref{eq:vprim} and \ref{eq:pbin} for the solar-type field star binary property distributions discussed above has been shown in Figure \ref{fig:cdf} as a dashed line for a primary mass of 1\,M$_\odot$. 

In practice we calculate the likelihood velocity distribution for binary-induced motions only for a single primary mass for a dense grid of possible radial velocity offsets. If the mass of an observed star is known, the likelihood distribution can be easily converted to this new primary mass by multiplying the grid of radial velocity offsets with $\left(\frac{m_1}{{\rm M}_{\odot}}\right)^{\frac{1}{3}}$ and dividing the computed probabilities by the same amount (keeping the normalization intact). If the period, mass ratio, or eccentricity distribution are known to change as a function of mass or some other observable (e.g. distance from the cluster center), we have to recalculate the likelihood function for binary orbital motions (eqs. \ref{eq:vprim} and \ref{eq:pbin}) for every observed star. 

\subsection{Fitting the parameters\label{sec:fit}}
Using equation \ref{eq:ptotal} we can combine the intrinsic velocity distribution predicted by the dynamical model (e.g. eq. \ref{eq:punder}) with the distribution of velocities due to binary orbital motions (eqs. \ref{eq:vprim} and \ref{eq:pbin}) to calculate the likelihood of observing a given velocity as a function of the binary fraction and any parameters describing the dynamical model of the intrinsic velocity distribution. In its most general form this can be written as $\mathcal{L}_i(v_i | \mathbf{x})$, where $v_i$ is the observed velocity, $\mathbf{x}$ is the set of free parameters, and $\mathcal{L}_i$ is the likelihood function, which will be different for every star, because it depends on the measurement error of the observation, the mass of the star, and any other observables included in the dynamical model. The total likelihood of reproducing all observed velocities is then given by $\mathcal{L}(\mathbf{v} | \mathbf{x}) = \prod_i \mathcal{L}_i(v_i | \mathbf{x})$. For a large number of observed stars this number will generally get too small for computers to compute (even for a well fitting model), so in practice we calculate the logarithm of the likelihood given by $\ln (\mathcal{L}(\mathbf{v} | \mathbf{x})) = \sum_i \ln (\mathcal{L}_i(v_i | \mathbf{x}))$. 

The likelihood distribution of reproducing the observation given the parameters ($\mathcal{L}(\mathbf{v} | \mathbf{x})$), which has been computed above, has to be converted to the probability distribution of the free parameters given the observations ($P(\mathbf{x} | \mathbf{v})$). Without any prior information on the probability distribution of the parameters, Bayesian inference states that $\mathcal{L}(\mathbf{v} | \mathbf{x})$ and $P(\mathbf{x} | \mathbf{v})$ have the same distribution, with the only difference being is that $P(\mathbf{x} | \mathbf{v})$ has been normalized
\begin{equation}
  P(\mathbf{x} | \mathbf{v}) = \frac{\mathcal{L}(\mathbf{v} | \mathbf{x})}{\int_{\mathbf{x}} \mathcal{L}(\mathbf{v} | \mathbf{x})\,{\rm d}\mathbf{x}} . \label{eq:bayes}
\end{equation}
Because the two distributions have the same shape, the set of parameters which maximize the likelihood of reproducing the observations will be the best-fit parameters.

As the number of fitted parameters increases, calculating the normalization in equation \ref{eq:bayes} becomes increasingly computationally expensive. An efficient way to explore the probability distribution of the parameters is through a Markov Chain Monte Carlo simulation. This procedure will take a constrained random walk through the parameter space, producing a chain of parameters, whose distribution follows the probability distribution. So the distribution of the values of a single parameter in this chain, will provide the probability distribution for this paramater, which will be automatically properly marginalized over all the other free parameters.

\section{Binary properties\label{sec:bin}}
Because there is a degeneracy between the effects of the period distribution, mass ratio distribution, eccentricity distribution and binary fraction on the observed single-epoch velocity distribution, it is not possible to fully constrain the binary properties using only a single epoch of radial velocity data. So a set of assumptions about these properties will have to be made to model the velocity distribution due to the orbital motions as described in Sec. \ref{sec:vdist}. Because the binary properties have been shown to significantly change with mass \citep{Burgasser07, Raghavan10, Chini12}, the choice of a period, mass ratio, and eccentricity distribution will generally depend on the mass range of the observed stars.

\begin{figure}
  \begin{center}
    \includegraphics[width=.5\textwidth]{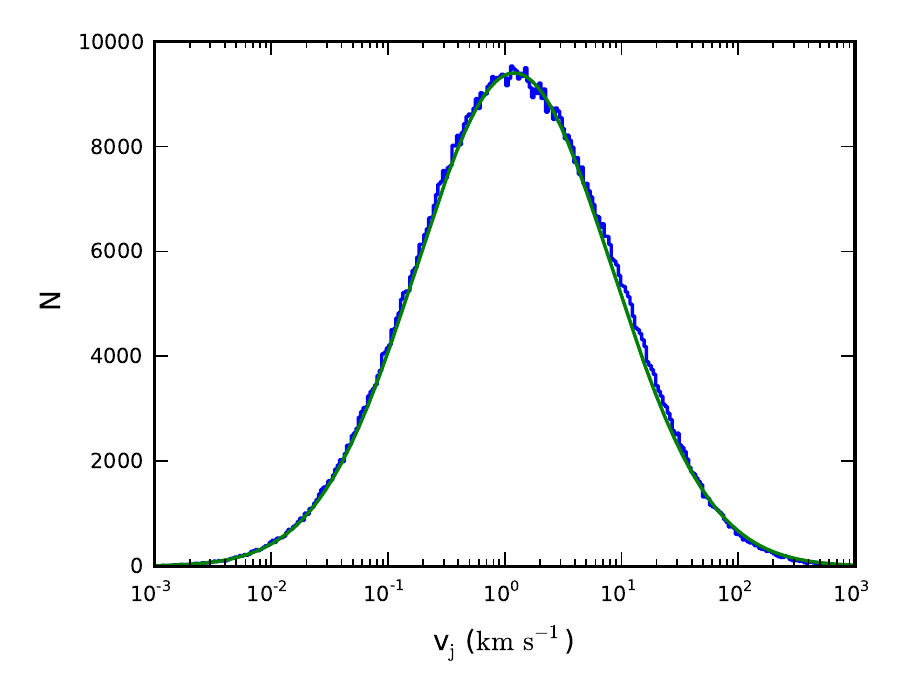}
		\caption{\label{fig:vbindist} The distribution of absolute velocities (i.e. not projected along the line of sight) due to binary orbital motions calculated for the period, mass ratio, and eccentricity distributions of the solar-type field stars \citep[see text;][]{Raghavan10, Reggiani11} calculated from equation \ref{eq:vprim} in blue and the best-fit log-normal in green. The log-normal has a width of $0.84 \log_{10}({\rm km\,s}^{-1})$ and a mean of $0.08 \log_{10}({\rm km\,s}^{-1})$.}
	\end{center}
\end{figure}

In many clusters the observed stars will tend to be solar-type due to either the evolution of the more massive stars in dwarf spheroidals or low-mass globular clusters or due to the lack of massive stars in low-mass local star-forming regions. So throughout this paper we will use the well observed binary properties for the solar-type field stars.  For the period distribution we use the log-normal distribution with a mean period of 5.03 and a dispersion of 2.28 in log$_{10}$\,days, which was found by \citet{Raghavan10}. For the secondary to primary mass ratio ($q$) we use the power-law from \citet{Reggiani11} of $\frac{{\rm d}N}{{\rm d}q} \sim q^{-0.5}$ for $0.1 < q < 1$. By setting the maximum $q$ to one we make the assumption that the observed (brightest) star in a binary will always be the most massive one. \citet{Raghavan10} found a lack of high eccentricities for binaries with low periods. We take this into account be drawing the eccentricities from a flat probability distribution between 0 and $e_{\rm max}$ with 
\begin{equation}
  e_{\rm max} = \frac{1}{2} [0.95 + \tanh(0.6 \log_{10} \frac{P}{\rm days} -1.7)],
\end{equation}
as proposed in \citet{Parker09a}. For these binary properties the distribution of the sizes of the three-dimensional velocity differences between the primary star and the center of mass of the binaries (calculated with eq. \ref{eq:vprim})  can be very well approximated by a log-normal with a mean of $0.08 \log_{10} {\rm km\,s}^{-1}$ and a dispersion of $0.84 \log_{10} {\rm km\,s}^{-1}$ (see Figure \ref{fig:vbindist}).

An important caveat with using the binary properties of the field population is that the solar-type binary properties might not be universal. The field star binary population might simply be a superposition of the binary properties in a large range of different star-forming regions with different binary properties. Furthermore we expect dynamical evolution to have altered the initial binary properties in older clusters \citep[e.g.][]{Marks11}.

Despite these caveats we show in Sec. \ref{sec:ngc188} that we can accurately describe the single-epoch velocity distribution of the dynamically old open cluster NGC 188 using the binary properties of the solar-type field stars. In Sec. \ref{sec:systematic} we will discuss the systematic offsets in the measured velocity dispersion and binary fraction induced, when the binary properties in the observed cluster do not match these assumptions.

\section{Accuracy of the procedure\label{sec:accuracy}}
Besides the statistical (random) uncertainty, which can be calculated through the Markov Chain Monte Carlo simulations, there might be an additional systematic error in the best-fit intrinsic velocity distribution if the assumed binary properties do not accurately describe the binaries in the cluster. Here we study the behavior of both the statistical and systematic uncertainty through a large suite of Monte Carlo simulations, where we create fake sets of radial velocity data and fit these with the procedure described in Sec. \ref{sec:method}. 

In the Monte Carlo simulations described in Sec. \ref{sec:random} we ensure that there is no systematic error by using the same binary properties to create the radial velocity data sets as we use to fit them, so we can focus purely on the statistical uncertainty and its dependence on the binary fraction, velocity dispersion, and number of observed stars. In Sec. \ref{sec:systematic} on the other hand we greatly increase the number of observed stars to minimize the statistical uncertainty, so that we can focus on the effect on the measured velocity dispersion if we use different binary properties to create the data sets as we use to fit them.

Throughout this section we will ignore the effect of the measurement errors of the observed radial velocities. For a normally distributed measurement error and a Gaussian intrinsic velocity distribution, the measurement error and the velocity dispersion are effectively indistinguishable (eq. \ref{eq:punder}). As long as the measurement error is sufficiently small that $\sigma_c^2 \approx \sigma_c^2 + \sigma_i^2$, the effect of the measurement error on the observed velocity distribution is indeed negligible. For a larger measurement error we can still use the results from this section, however the velocity dispersion should be replaced with the quadratic sum of the velocity dispersion and the measurement error. In this case the measurement error will have to be well quantified in order to retrieve the velocity dispersion.

\subsection{Statistical uncertainty\label{sec:random}}
\begin{figure*}
	\begin{center}
 		\includegraphics[width=\textwidth]{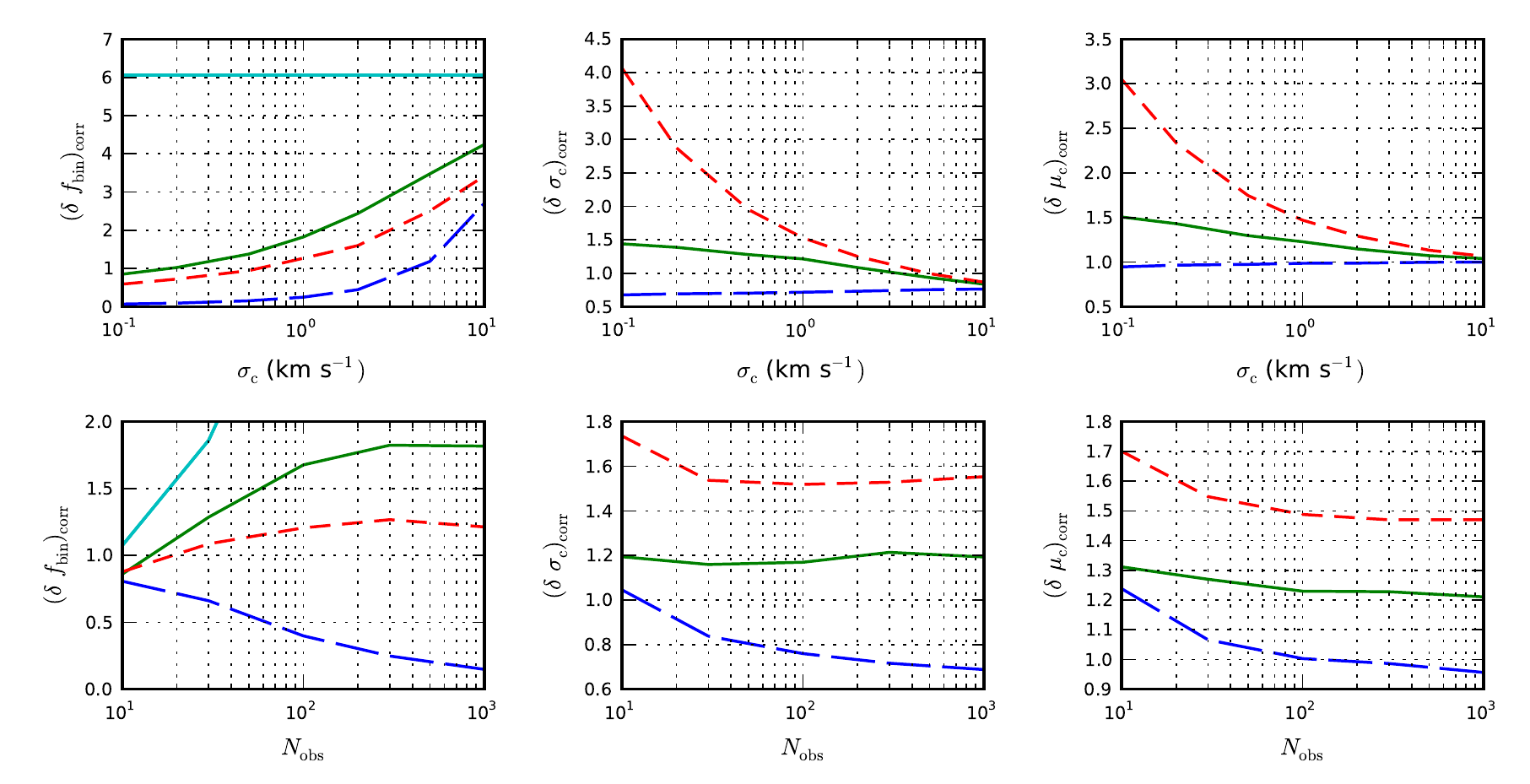}
		\caption{\label{fig:accuracy} From left to right the second-order corrections of the uncertainties of the binary fraction $(\delta f_{\rm bin})_{\rm corr}$, the velocity dispersion $(\delta \sigma_{\rm c})_{\rm corr}$, and the mean velocity $(\delta \mu_{\rm c})_{\rm corr}$. The top panels show the dependency of these corrections on the velocity dispersion of the cluster ($\sigma_c$) and the lower panels show their dependency the sample size ($N_{\rm obs}$). These corrections have been computed for the  a binary fraction of 0\% (blue), 50\% (green), and 100\% (red). They can be used to calculate the one-sigma uncertainties using equations \ref{eq:acc_fbin} - \ref{eq:acc_vmean}.  This allows for the estimation of the accuracy with which the intrinsic velocity distribution can be measured before any radial velocity data has been taken, as long as an estimate for the velocity dispersion and binary fraction are available. The cyan lines in the left panels show the lines with $\delta f_{\rm bin} = 0.34$, which is the expected uncertainty if the binary fraction is unconstrained (i.e. a flat probability distribution between 0 and 1).}
	\end{center}
\end{figure*}
After the velocities have been measured the statistical uncertainty can always be calculated through Markov Chain Monte Carlo simulations. When preparing for observations it is often useful to have an estimate of the accuracies that can be reached based only on the number of stars that will be observed and the expected properties of the cluster prior to any radial velocity measurements.

With this goal in mind we ran Monte Carlo simulations on a large number of fake sets of radial velocities with solar type binary properties (Sec. \ref{sec:bin}) with varying binary fractions, velocity dispersions, and sample sizes. These data sets were created for stars with random primary masses between 0.1 and 1\,M$_\odot$ following the \citet{Chabrier05} IMF, which is an update of the IMF presented in \citet{Chabrier03}. For every star we calculated a random radial velocity from a normal distribution with the chosen velocity dispersion (eq. \ref{eq:punder}). Every star has a probability set by the binary fraction to be part of a binary system, in which case an additional velocity due to the binary orbital motion is added. The binary properties are randomly chosen from the period, mass ratio, eccentricity and phase distributions and the binary is randomly orientated with respect to the line of sight. The radial velocity dataset created in this way is then fitted using the procedure described in Sec. \ref{sec:method}. Because we used the same assumptions about the intrinsic velocity distribution and the binary properties to create and to fit the radial velocities, there are no systematic offsets in the measured velocity dispersion, mean velocity, and binary fraction with respect to the input velocity dispersion, mean velocity, and binary fraction. For every simulation the uncertainties on the best-fit parameters were calculated through Markov Chain Monte Carlo simulations.

To first order (i.e. ignoring binaries) the uncertainty in the mean velocity is given by the standard deviation of the mean$\delta \mu_{\rm c} = \frac{\sigma_{\rm c}}{\sqrt{N_{\rm obs}}}$, where $\sigma_{\rm c}$ is the velocity dispersion and $N_{\rm obs}$ is the sample size. Similarly we expect the velocity dispersion to go as $\delta \sigma_{\rm c} \sim \frac{\sigma_{\rm c}}{\sqrt{N_{\rm obs}}}$ and finally the uncertainty in the binary fraction to go with $\delta f_{\rm bin} \sim \frac{1}{\sqrt{N_{\rm obs}}}$.

Using these first-order effects we can write the uncertainty in the binary fraction, velocity dispersion, and mean velocity as
\begin{subequations}
\begin{align}
 \delta f_{\rm bin} &= \frac{(\delta f_{\rm bin})_{\rm corr}}{\sqrt{N_{\rm obs}}}, \label{eq:acc_fbin}\\
 \delta \sigma_{\rm c} &= \frac{(\delta \sigma_{\rm c})_{\rm corr} \ \sigma_{\rm c}}{\sqrt{N_{\rm obs}}}, \label{eq:acc_vdisp}\\
 \delta \mu_{\rm c} &= \frac{(\delta \mu_{\rm c})_{\rm corr} \ \sigma_{\rm c}}{\sqrt{N_{\rm obs}}}, \label{eq:acc_vmean}
\end{align}
\end{subequations}
where the corrections for the second-order effects due to binaries are given by $(\delta f_{\rm bin})_{\rm corr}$, $(\delta \sigma_{\rm c})_{\rm corr}$, and $(\delta \mu_{\rm c})_{\rm corr}$. Figure \ref{fig:accuracy} shows the dependence of these quantities on the velocity dispersion of the cluster (for $N_{\rm obs} = 300$) and the number of observed stars (for $\sigma_c=1$\,km\,s$^{-1}$) for a binary fraction of 0\% (blue, long dashes), 50\% (green, solid line), and 100\% (red, short dashes). 

The bottom panels in Figure \ref{fig:accuracy} show on the number of observed stars ($N_{\rm obs}$) for a fixed velocity dispersion of 1\,km\,s$^{-1}$. We find that the second-order corrections on the velocity dispersion and mean velocity becomes independent of $N_{\rm obs}$ for $N_{\rm obs}>50$. This means that the accuracy of the velocity dispersion and mean velocity can be well approximated to go with $1/\sqrt{N_{\rm obs}}$. This dependence becomes unreliable for $N_{\rm obs} < 50$, because the measurement errors themselves vary greatly for small $N_{\rm obs}$ and the uncertainty in the velocity dispersion becomes strongly non-Gaussian. 

The uncertainty in the binary fraction (bottom left panel) shows a similar behavior, as long as the binary fraction is not too low. For a binary fraction of 50\% (green) and velocity dispersion of 1\,km\,s$^{-1}$ the binary fraction becomes effectively unconstrained between 0 and 100\% for $N_{\rm obs} < 100$. For a binary fraction of 0\% we find that $(\delta f_{\rm bin})_{\rm corr}$ declines as the sample size increases, implying that the measurement error in the best-fit binary fraction decrease more steeply than $1/\sqrt{N_{\rm obs}}$ over the full range of tested $N_{\rm obs}$.

Because the second-order corrections $(\delta f_{\rm bin})_{\rm corr}$, $(\delta \sigma_{\rm c})_{\rm corr}$, and $(\delta \mu_{\rm c})_{\rm corr}$ are constant over a broad range of $N_{\rm obs}$ it is possible to calculate the accuracy expected for future measurements from the top panels in Figure \ref{fig:accuracy} using eqs \ref{eq:acc_fbin}-\ref{eq:acc_vmean} not only for  $N_{\rm obs}=300$ for which the figure was calculated, but for any $N_{\rm obs}>50$ given a prior estimate of the velocity dispersion and binary fraction. To fit more complicated dynamical models with more free parameters more stars will of course need to be observed to reach the same accuracy.

For smaller sample sizes these estimates should be taken with caution. For small $N_{\rm obs}$ the uncertainties themselves become statistical quantities, which might randomly be much larger (or smaller) than expected. Furthermore the probability distribution of the parameters become non-Gaussian for small sample sizes.

Keeping in mind these caveats eqs \ref{eq:acc_fbin} - \ref{eq:acc_vmean} and the top panels in Figure \ref{fig:accuracy} can be used to estimate the sample size needed to reach a certain accuracy on the observed velocity dispersion. For example if we consider a local star-forming regions with an expected velocity dispersion of about 2 km\,s$^{-1}$ and solar-type field star binary properties with a binary fraction of roughly 50\%, we find from the upper center panel in Figure \ref{fig:accuracy} $(\delta \sigma_{\rm c})_{\rm corr} = 1.1$. To reach an accuracy of 10\% on the velocity dispersion (i.e. $\frac{\delta \sigma_{\rm c}}{\sigma_{\rm c}} = 0.1$), we find from eq. \ref{eq:acc_vdisp} that $N_{\rm obs} = \left(\frac{1.1}{0.1}\right)^2=121$ stars would need to be observed to reach the desired accuracy.

The calculations above were done assuming that the measurement uncertainty is significantly smaller than the velocity dispersion, so that the measurement uncertainty has a negligible effect on the observed velocity dispersion (effectively replacing $\sigma_c^2 + \sigma_i^2 \approx \sigma_c^2$ in eq. \ref{eq:punder}). Actual observations will sometimes include stars for which the measurement uncertainty is comparable or even larger than the velocity dispersion. These velocities will have a broader likelihood function with the width set by the quadratic sum of the measurement uncertainty and the velocity dispersion (eq. \ref{eq:punder}). As seen in Figure \ref{fig:accuracy} and eqs. \ref{eq:acc_fbin} - \ref{eq:acc_vmean} the uncertainty in observed parameters increases for broader distributions, so these measurements will contribute less to the accuracy with which the intrinsic velocity dispersion and binary fraction are measured. However, as the measurement uncertainties are included in the model, these observations can be safely kept in the dataset as long as the measurement uncertainty is well measured. If the measurement uncertainty is only a rough estimate it might be better to remove the observations with the largest measurement uncertainties from the dataset, as an overestimate of the measurement error might cause the velocity dispersion to be systematically underestimated and vice versa.

\subsection{The systematic uncertainty due to inaccurate binary assumptions\label{sec:systematic}}

\begin{figure}
	\begin{center}
 		\includegraphics[width=.5\textwidth]{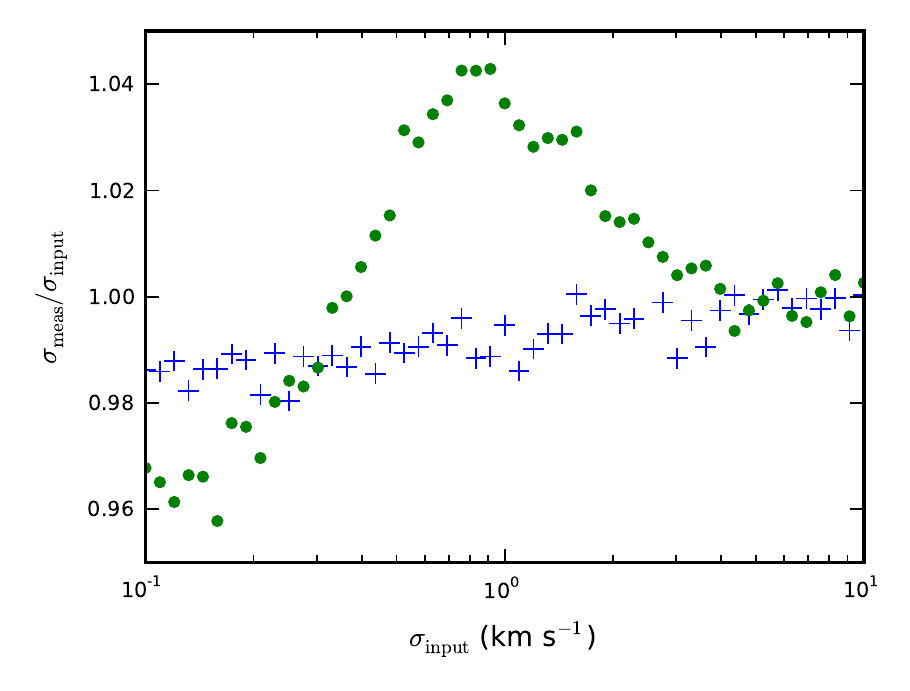}
		\caption{\label{fig:syst} The ratio between the measured and input velocity dispersion resulting from fits of artificial datasets with varying input velocity dispersions. These datasets were fitted assuming the binary properties of solar-type field star, but were created using different binary period and mass ratio distributions. For the blue '+'-signs the binary population used to create the velocity distribution follows the same field solar-type binary properties used to fit the distribution, except that the mass ratio distribution is flat rather than falling (see Sec. \ref{sec:qflat}). For the green dots the massive stars follow the solar-type binary properties, while the binary properties of the low mass stars are set by the most likely binary fraction, period distribution, and mass ratio distribution observed for very low mass binaries from \citet{Burgasser07} (see Sec. \ref{sec:mdep}).}
	\end{center}
\end{figure}

To fit an observed velocity distribution, we have to model the binary orbital motions. This requires assumptions for the binary period, mass ratio, and eccentricity distributions. If our assumed distributions do not match the actual binary properties in the observed cluster, there will be a systematic error in the measured velocity dispersion. 

Here we investigate the size of this systematic error by running a set of Monte Carlo simulations. In these simulations we create artificial radial velocity datasets for a given input velocity dispersion and a set of binary properties, which are different than those described for the solar-type field stars discussed in Sec. \ref{sec:vdist}. We fit these artificial radial velocity data sets assuming the binary period, mass ratio, and eccentricity distribution from the solar-type field stars discussed in Sec. \ref{sec:vdist}. We then compare the measured velocity dispersion with the input velocity dispersion used to build the artificial dataset and check for a systematic offset between the two.

The measured velocity dispersion will differ from the input velocity dispersion with a random offset due to the statistical noise discussed in Sec. \ref{sec:random} as well as a systematic offset due to the discrepancy between the binary properties used to create the artificial dataset and the binary properties used to fit the dataset. Because we are interested here in studying the systematic offset, we minimize the statistical noise by building a very large dataset of $10^5$ radial velocities. Even with $10^5$ observed stars there will still be some random scatter in the measured velocity dispersion and binary fraction.

The procedure to build the artificial dataset is the same as in Sec. \ref{sec:random}. We briefly summarize it again. We draw the masses of the `observed' stars following the \citet{Chabrier05} IMF between 0.1 and 1\,M$_\odot$. For every star a radial velocity is then drawn from a Gaussian distribution with a given input velocity dispersion. For the randomly assigned binaries a velocity offset due to the binary orbital motions is added.

\subsubsection{Flat mass ratio distribution\label{sec:qflat}}
To characterize the distribution of binary orbital motions we assumed a mass ratio distribution of $\frac{{\rm d}N}{{\rm d}q} \sim q^{-0.5}$\citep{Reggiani11}. However, \citet{Reggiani11} found an uncertainty on this power-law slope of $\frac{{\rm d}N}{{\rm d}q} \sim q^{-0.5 \pm 0.29}$. This implies that the observed mass ratio distribution is still consistent with being flat at the two sigma level. If the mass ratio distribution were in fact flat, this would cause a systematic offset in the measured velocity dispersion given our assumption of a power-law mass ratio distribution.

To explore this effect we ran a set of Monte Carlo simulations, where we create an artificial 'observed' radial velocity dataset assuming a flat mass ratio distributions between $q=0.1$ and $q=1$ for several input velocity dispersions ranging from 0.1 to 10\,km\,s$^{-1}$. To create these datasets we use the period and eccentricity distributions of the solar-type field stars and fix the binary fraction to the solar-type field star binary fraction \citep[46\%:][]{Raghavan10}. These artificial datasets are then fitted assuming the (correct) solar-type field star period and eccentricity distributions, but with the (here incorrect) power-law slope for the mass ratio distribution.  The velocity dispersion measured in this way have been plotted as blue '+'-signs in Figure \ref{fig:syst} for values of the input velocity dispersion between 0.1 and 10\,km\,s$^{-1}$.

The flat mass ratio distribution has a higher fraction of near equal-mass binaries, so every individual binary has a larger probability of creating a significant velocity offset than for the power law mass ratio distribution. So to reproduce the number of high velocity outliers in the artificial data set created with the flat mass ratio distribution, the fitted binary fraction will be higher than 46\% for the power-law mass ratio distribution. We find that the best-fit binary fraction is indeed higher than 46\% and increases from 48\% for an input velocity dispersion of 0.1\,km\,s$^{-1}$ to 55\% for an input velocity dispersion of 10\,km\,s$^{-1}$.

The effect of the flat mass ratio distribution is much smaller on the measured velocity dispersion than on the binary fraction. Figure \ref{fig:syst} shows the ratio of the measured and input velocity dispersion as blue '+'-signs. If the assumed binary properties would perfectly match those of the observed cluster this would always be one (except for some random scatter due to the statistical uncertainty discussed in Sec. \ref{sec:random}). We see that the difference between the flat and power-law slope for the mass ratio distribution causes an offset of $<$1\% to almost 2\%. This offset increases for lower input velocity dispersion, because the effect of binary orbital motions become more important for narrower velocity distributions.

\subsubsection{Mass-dependent binary properties\label{sec:mdep}}
We also consider the case, where the low-mass binaries (primary mass between 0.1 and 0.2 $M_\odot$) in our artificial cluster do not have the same properties as the solar-type field stars, but rather the period distribution, mass ratio distribution, and binary fraction found by \citet{Burgasser07} for very low mass binaries. The binary properties for these stars are still highly uncertain, but \citet{Burgasser07} found that the observed binary properties for binaries with primary masses lower than 0.1 $M_\odot$ could be fitted by a log-normal period distribution with a mean at 7 AU and a width of 0.24\,log(AU), a power-law mass ratio distribution with $\frac{{\rm d}N}{{\rm d}q} \sim q^{4.8}$ (for $q < 0.6$) and a binary fraction of 22\%. This binary fraction is much lower than that observed for solar-type stars in the field, although the effect per binary on the velocity distribution is much larger, because they tend to be on closer orbits and tend to be equal-mass.

In our simulations the masses range from 0.1 to 1\,M$_\odot$. Although recent results \citep{Janson12} suggest a smooth transition between the binary properties of the very low mass and solar-type stars, we will use a sharp transition between these regimes for this numerical test. For the low mass stars ($< 0.2\,{\rm M}_\odot$, $\sim$34\% of all stars) we use the semi-major axis log-normal distribution, the mass ratio power-law distribution (for $q$ between 0.1 and 1), and the binary fraction found by \citet{Burgasser07}. For the higher mass stars we use the binary properties observed for the solar-type field stars (Sec. \ref{sec:bin}) with a binary fraction of 46\%. We once again fit these radial velocities under the (inaccurate) assumption that all stars still have the same binary fraction and follow the solar-type field binary period, mass ratio, and eccentricity distributions. These simulations are repeated for a broad range of input velocity dispersions. 

The measured velocity dispersion is shown in Figure \ref{fig:syst} as green dots. Although the assumed solar-type field binary properties are inaccurate for a third of the 'observed' stars, the offset between the measured and input velocity dispersion is still at most 4\%.

These two examples (Sec. \ref{sec:qflat} and \ref{sec:mdep}) are meant to illustrate the effect of incorrect binary assumptions on the measured intrinsic velocity distribution. For reasonable differences between the true and assumed binary properties, the effect on the intrinsic velocity dispersion turns out to be fairly small. For the cases discussed above only a few percent. It would require observations of thousands of stars, before this systematic offset becomes significant compared to the statistical error. However, if the binary properties are very uncertain, it might still be useful to check how much the intrinsic velocity distribution might vary for several reasonable assumptions about the binary properties \citep{Odenkirchen02}.

\section{Case study: NGC 188\label{sec:ngc188}}
As a test for the procedure described above, we will here consider the case of the old \citep[$7 \pm 0.5$\,Gyr;][]{Sarajedini99} open cluster NGC 188. The dynamical state of the cluster was extensively studied by \citet{Geller08,Geller09} and \citet{Geller11}. Based on 9166 radial velocity measurements of 1108 stars in the direction of this cluster, \citet{Geller08} identified a large number of stars with significant radial velocity variations. Based on the remaining (seemingly single) 640 stars, they derive a velocity dispersion of $0.64 \pm 0.04\ {\rm km\,s}^{-1}$. This value is still inflated by unidentified binaries. \citet{Geller11} found that these unidentified binaries would inflate an intrinsic velocity dispersion of $0.49 ^{+0.07}_{-0.08}\ {\rm km\,s}^{-1}$ to the observed $0.64 \pm 0.04\ {\rm km\,s}^{-1}$. In this section we will show that we can reproduce their intrinsic radial velocity distribution using only a single epoch of their data. 

We estimate the masses of the observed stars to be given by the closest point on the 7 Gyr Padova isochrone \citep{Marigo08} with ${\rm E}({\rm B}-{\rm V})=0.09$ and DM$=11.44$ \citep[following the photometric fit of][]{Sarajedini99}. The masses of the observed stars estimated in this manner vary between 0.8 and 1.1\,M$_\odot$. The sample of observed stars contains a significant contamination from fore- and background stars. For these stars we do not expect the computed masses to be accurate. Fortunately in our model the masses only affect the binary orbital motions, which have a negligible effect on the very broad velocity distribution of the fore- and background stars.

These fore- and background stars will affect the intrinsic velocity distribution. Following \citet{Geller08} we model the radial velocity distribution of the field star population with a Gaussian. Choosing the same distribution to fit the field stars allows us to make direct comparison with the results from \citet{Geller08}. We can add this Gaussian to the model for the intrinsic velocity distribution (eq. \ref{eq:punder}) to get
\begin{equation}
\begin{split}
\mathcal{L}_{\rm dyn,\ i}(v_{\rm dyn})= &f_{\rm c} \frac{1}{\sqrt{2 \pi (\sigma_{\rm i}^2+\sigma_{\rm c}^2)}} \exp\left(-\frac{(v_{\rm dyn}-\mu_c)^2}{2 (\sigma_{\rm i}^2+\sigma_{\rm c}^2)}\right) +\\
 &(1-f_{\rm c})\frac{1}{\sqrt{2 \pi (\sigma_{\rm i}^2+\sigma_{\rm f}^2)}} \exp\left(-\frac{(v_{\rm dyn}-\mu_{\rm f})^2}{2 (\sigma_{\rm i}^2+\sigma_{\rm f}^2)}\right),
\end{split} \label{eq:punder_cont}
\end{equation}
where $f_{\rm c}$ is the probability of an observed stars to be a cluster member (i.e. the fraction of cluster members) and $\sigma_{\rm f}$ and $\mu_{\rm f}$ are respectively the velocity dispersion and mean velocity of the non-members. These three parameters will be fitted in addition to the binary fraction ($f_{\rm bin}$), the cluster velocity dispersion ($\sigma_{\rm c}$), and the mean velocity ($\mu_{\rm c}$) of the cluster. 

\begin{figure}
	\begin{center}
 		\includegraphics[width=.5\textwidth]{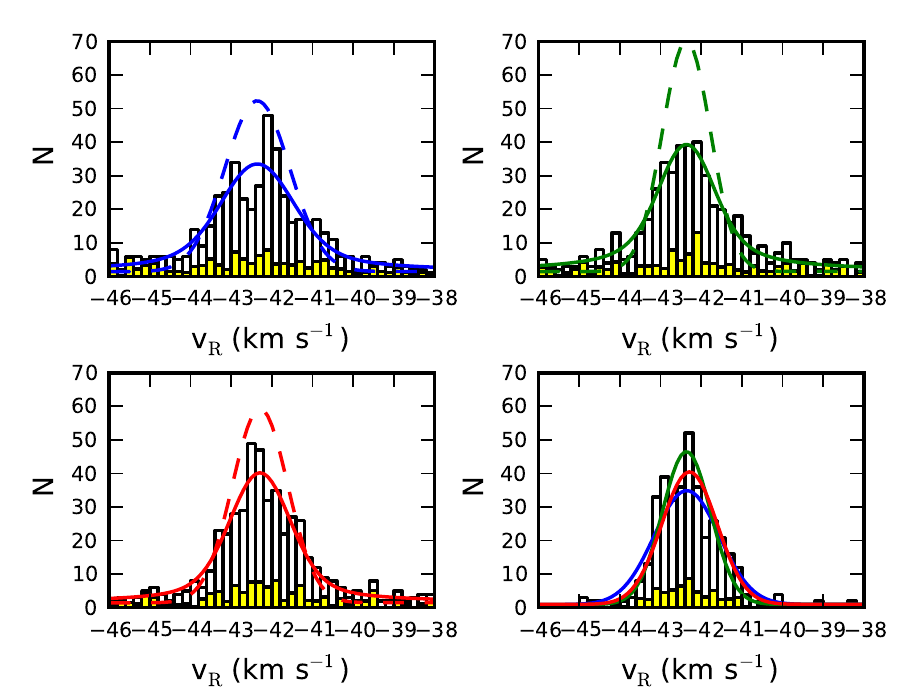}
		\caption{\label{fig:ngc188_dist} Histograms of the three single-epoch radial velocity distribution from NGC 188 we consider in the top left, top right and bottom left panels, with the probability distribution of the best fitting model overplotted as the solid line for a star of 1\,M$_\odot$ and a measurement error of 0.4\,km\,s$^{-1}$. The intrinsic velocity distributions (eq \ref{eq:punder_cont}) are plotted as dashed lines. In the bottom right the histogram of the velocities of the single stars identified in \citet{Geller08} is plotted. The intrinsic velocity distributions (renormalized to the smaller population) computed from the three data sets have been overplotted. All three of these intrinsic velocity distributions are consistent with the velocity distribution of the single stars from \citet{Geller08} according to a KS-test (see text). In all panels the fraction of every bin expected to consist of non-members \citep{Platais03} have been made yellow. The color coding of the lines corresponds to the data set used to fit the model. The first observed epoch has been shown in blue, the second epoch in green, and the last epoch in red.}
	\end{center}
\end{figure}

Because multiple epochs of data are available for nearly all of the 1108 stars we are able to run the procedure on three single-epoch radial velocity data sets with only minimal duplication between the data sets. The radial velocities and measurement errors were taken from \citet{Geller09}. 

The first three panels in Figure \ref{fig:ngc188_dist} show the histograms of the radial velocities in these three data sets with the velocity distribution for the best-fit model overplotted as solid lines. In principle there is no single best-fit distribution, because the velocity distribution depends on the mass of the star and the measurement error. The distributions plotted in Figure \ref{fig:ngc188_dist} are for a star of 1\,M$_\odot$ and a measurement error of 0.4\,km\,s$^{-1}$, which is typical for the vast majority of observed stars. We note that the model has been fitted to the individual observed velocities and is independent of the bin size chosen in Figure \ref{fig:ngc188_dist}. 

The velocity distribution of single stars identified in the multi-epoch data by \citet{Geller08} is shown in the bottom right panel in Figure \ref{fig:ngc188_dist}, where we have overplotted the intrinsic velocity distributions found for the first data set (blue), the second data set (green), and the third data set (red). We check whether the observed radial velocity distribution of the single stars is consistent with being drawn from the predicted intrinsic velocity distributions from the single epoch datasets using the Kolmogrov-Smirnov test. The KS test computes the largest deviation between the measured and predicted cumulative distributions and calculates the probability that this deviation or a larger one can be created by chance when drawing random samples from the predicted distribution. If this probability is small, the hypothesis that the measured and predicted distributions are the same can be rejected.

Before we can compare the observed single star velocity distribution with the intrinsic velocity distribution predicted from the single-epoch data, we have to correct for a selection effect. Because priority was given during the observations to proper motion members, the fraction of cluster members in the subset of observed stars with at least three epoch of observations over a baseline of one year is higher than in all observed stars. Because \citet{Geller08} require stars to have been observed for at least three epochs over at least one year before classifying them as single stars, the fraction of members among the stars identified as single is also higher. Based on the proper motions membership probabilities \citep{Platais03} we find the fraction of members among stars classified as single to be 53\%, significantly higher than the 44\% among all stars.

After correcting for this selection effect by increasing $f_{\rm c}$ to $53\%$ the KS test finds that the observed velocity distribution of the seemingly single stars is consistent with being drawn from the predicted intrinsic velocity distribution. The probabilities of drawing a distribution with at least the same offset in the cumulative distribution by chance is 17\% for the first data set, 29\% for the second data set, and 2\% for the last data set. We note that even in the unlikely case that the assumptions going into the model (i.e. Gaussian distribution, field solar-like binary properties) are perfect descriptions of NGC 188, the distributions are still expected to be slightly different, because the observed distribution of stars identified as single might still be inflated by unidentified binaries. 

\begin{figure}
	\begin{center}
 		\includegraphics[width=.5\textwidth]{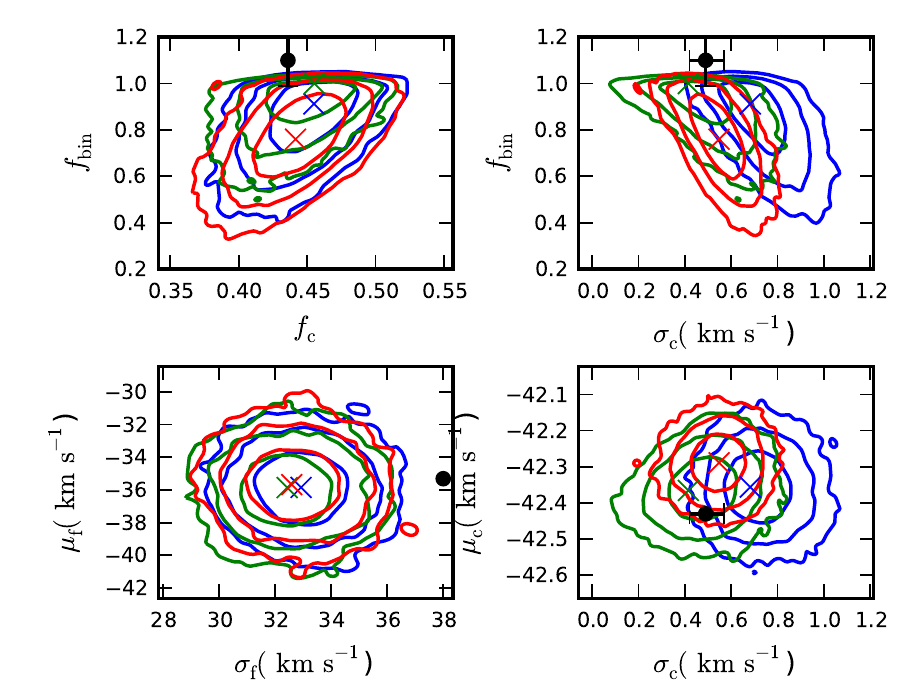}
		\caption{\label{fig:ngc188_cont} Posterior probability distributions of the free parameters in the model, enclosing 68.2\%, 95.4\%, and 99.7\% (i.e. 1, 2, and 3 $\sigma$) of the total probability. The six free parameters are the binary fraction ($f_{\rm bin}$), the fraction of cluster members ($f_{\rm c}$), the cluster velocity dispersion ($\sigma_{\rm c}$), the cluster mean velocity ($\mu_{\rm  c}$), the field velocity dispersion ($\sigma_{\rm f}$) and the field mean velocity ($\mu_{\rm f}$). The different colors refer to the three different single-epoch radial velocity distributions with the same color coding as in Figure \ref{fig:ngc188_dist}. The literature values for these parameters, derived from a proper motion analysis \citep{Platais03} and the multi-epoch radial velocity data \citep{Geller08, Geller11} have been plotted as black dots. If the error bars are given, they have been plotted.}
	\end{center}
\end{figure}

In Figure \ref{fig:ngc188_cont} we show the probability distributions for the six parameters we fit in our model. For comparison the literature values for these parameters from \citet{Platais03, Geller08, Geller11} have been overplotted. Taking the average of all membership probabilities for all observed stars \citep{Platais03}, we find that 44\% of observed stars should be members. This is fully consistent with the results we find for the three individual data sets (top left panel of Figure \ref{fig:ngc188_cont}).

For all three datasets we find a binary fraction consistent with being 100\% (Fig. \ref{fig:ngc188_cont}). This high binary fraction should be interpreted with great caution, as we are only sensitive to the close binaries in NGC 188, which affect the observed velocity distribution. If a larger fraction of all binaries in NGC 188 is close compared to the solar-type field binaries, the total binary fraction might be significantly lower. Such a deviation from the binary properties of the solar-type field stars might be caused by dynamical evolution, which decreases the period of the hard binaries \citep[$P < 10^6$ days in the case of NGC 188,][]{Geller11}. Evidence for significant dynamical evolution in NGC 188 was found by \citet{von-Hippel98} and \citet{Bonatto05}.

Irrespective of whether there are wide orbit counterparts to the identified spectroscopic binaries, we can compare the binary fraction found to the spectroscopic binary fraction from the multi-epoch data. \citet{Geller11} found that $29\pm3\%$ of observed stars in NGC 188 are binaries with periods smaller than $10^4$ days and $q>0.2$, which makes these type of binaries $2.4\pm0.2$ times as frequent in NGC 188 as in the field. Under our assumptions that the binaries in NGC 188 have the same properties as those in the field, the total binary fraction would then also be $2.4\pm0.2$ times as high. This would lead to the unphysical conclusion that $110 \pm 10$\% of stars in NGC 188 are binaries, which is consistent with the very high binary fractions found from the single-epoch data. The estimate of a total binary fraction above the 100\% implies that the ratio of spectroscopic binaries to wider binaries is higher in NGC 188 than in the field.

We finally compare the results of our Gaussian fits. \citet{Geller08} measured a mean velocity of -35.31 km\,s$^{-1}$ and a dispersion of 38.01 km\,s$^{-1}$ for the non-members. The cluster members have a mean velocity of -42.43 km\,s$^{-1}$ \citep{Geller08} and a velocity dispersion of $0.49 ^{+0.07}_{-0.08}\ {\rm km\,s}^{-1}$ \citep{Geller11}. All of the mean velocities and velocity dispersions measured for our three single-epoch datasets, except possibly the velocity dispersion of the field population for which \citet{Geller08} do not give an uncertainty, are consistent with the literature values within one to two sigma.

\section{Discussion\label{sec:disc}}
Above we have shown that the presented likelihood procedure can accurately describe the intrinsic velocity distribution of NGC 188 and find an intrinsic velocity dispersion consistent with that the $0.49 ^{+0.07}_{-0.08}\ {\rm km\,s}^{-1}$ from multi-epoch data \citep{Geller11}. This dispersion is accurately found, even though NGC 188 has a large spectroscopic binary fraction and there is a large contamination of fore- and background stars.

In our analysis of NGC 188 we have attempted to use as little prior information about the cluster as possible in order to illustrate the accuracy of the procedure based only on the observed single-epoch radial velocity and the mass estimated of the stars. The accuracy of the procedure could of course be improved by taking constraints from other sources into account. For example one could use the spectroscopic binary properties which \citet{Geller11} found for NGC 188 instead of the solar-type field binary properties. Furthermore one could set the probability for a star to be a cluster member $f_{\rm c}$ to the membership probabilities derived from the proper motion analysis of \citet{Platais03}. Finally one could use a model for the Milky Way \citep[e.g.][]{Robin03} to predict the velocity distribution of the fore- and background star instead of fitting a Guassian to this distribution.

Although we have tested the procedure here on an open cluster, it should be applicable to a broad range of stellar systems where the velocity distribution is significantly inflated due binary orbital motions, such as faint Milky Way companion dwarf galaxies, young massive clusters, local star-forming regions, and some low-mass globular clusters. As long as there is a dynamical model for the velocity distribution of single stars and the center of masses of binaries and a rough approximation of the binary properties, a single epoch of radial velocity data can be used to constrain the dynamical state of the cluster. If enough velocity data is available it can serve as a test on the assumed binary properties for the observed cluster.

\subsection{Extension to multiple epochs}
Up till now we have only discussed the application of this procedure to single-epoch data, because that is were the gain of using this procedure is by far the largest. However, the method can also be applied to multi-epoch data. In theory the ideal procedure would be to compute for every star the probability distribution of possible center of mass velocities based on the full ensemble of binary orbital solutions, which are consistent with the data. This can then be used to constrain the intrinsic velocity distribution. However, if there are only a small number of epochs per star, calculating the full ensemble of binary orbital solutions would be computationally very expensive.

An alternative procedure would be to only use those stars, whose radial velocity variations are consistent with them being single \citep[e.g.][]{Geller08, Geller10}. The observed velocity distribution of these stars would still be affected by any undetected binaries. The velocity offsets due to these binaries can be modeled by drawing a large sample of binaries with random periods, mass ratios, eccentricities, phases, and orientations. Any binaries which would have been detected given the timing of the observations and the measurement errors are removed from this sample \citep[e.g.][]{Geller10} . The distribution of velocity offsets caused by the remaining binaries, which would not be detected in the multi-epoch observations, is then set to the distribution of velocities due to binary orbital motions ($\mathcal{L}_{\rm bin ,\ i} (v_{\rm bin})$). This distribution can be used as input to fit the velocity distribution corrected for any undetected binaries, using exactly the same procedure as described in Sec. \ref{sec:method}. 

\subsection{Limitations}
The power of this procedure lies in the difference between the Gaussian intrinsic velocity distribution assumed and the high velocity outliers caused by the binary orbital motions. This implies that if a dynamical model is fitted, where the intrinsic velocity distribution is more similar to the distribution caused by binary orbital motions (i.e. has stronger wings), the accuracy of the procedure would decrease. An example would be a model, which includes the possibility of stars ejected at a high velocity. To distinguish these ejected stars from velocity outliers caused by binary orbital motions, multiple epochs of radial velocity data will still be required. 

We have a similar problem in our analysis of the radial velocity data of NGC 188 (Sec. \ref{sec:ngc188}) where the nature of the high-velocity outliers is unclear. They could either be members, whose velocity is offset due to its orbital motion, or they could be non-members. Due to this confusion the accuracy with which the binary fraction could be measured, was much lower as expected for such a large set of radial velocity data. From the top left panel of Figure \ref{fig:accuracy} we find that $(\delta f_{\rm bin})_{\rm corr}$ for a cluster with a binary fraction of roughly 100\% and a velocity dispersion of 0.5\,km\,s$^{-1}$ is about 0.6. For the $N \approx 500$ members in NGC 188 equation \ref{eq:acc_fbin} gives an accuracy of 3\%. This is much lower than the actual precision with which we could measure the binary fraction of $\sim 15\%$.

In addition to distinguishing between velocity outliers due to binary orbital motions and those caused by ejections or non-members, multi-epoch data has the advantage that it can be used to constrain the binary properties of the cluster. In principle the distribution of the high-velocity tail in the observed velocity distribution puts some constraints on the binary properties. If enough radial velocity data is available, this might be sufficient to reject a given set of binary assumptions. However the degeneracy between the effects of the period, mass ratio, and eccentricity distributions and the binary fraction on this high-velocity tail means that these distributions can not actually be fitted. 

\section{Summary\label{sec:summary}}
We investigate here a maximum likelihood method to simultaneously fit an intrinsic velocity distribution with the velocity offsets induced due to binary orbital motions to a single epoch of observed radial velocities. The procedure was pioneered seperately by \citet{Odenkirchen02} and \citet{Kleyna02}, but has not been used since. We show it can be used on other types of clusters, where it enables a proper treatment of the effect of binary orbital motions on the radial velocities observed in surveys of single-epoch high-resolution spectra.

This procedure requires assumptions about the period, mass ratio, and eccentricity distribution for the binaries in the observed stars, although we show that reasonable uncertainties in the assumed binary properties do not significantly affect the intrinsic velocity distribution we find (Sec. \ref{sec:systematic}). 

Assuming a binary period, mass ratio, and eccentricity distribution found for solar-type stars in the field \citep{Raghavan10} the precision with which the binary fraction, velocity dispersion, and mean velocity can be measured as a function of the number of observed stars, the velocity dispersion, and the binary fraction can be determined using eqs. \ref{eq:acc_fbin} - \ref{eq:acc_vmean} and Figure \ref{fig:accuracy}. These results can be used to get a prior estimate of the number of stars that should be observed to reach a given accuracy. Although we calculate the accuracy with which the velocity dispersion can be measured, we emphasize that the procedure can be trivially expanded to test complicated dynamical models, where for example the velocity dispersion and the mean velocity depend on the position of the observed star in the cluster and/or its mass.

As a test case we apply this procedure to three different sets of single epoch radial velocity data from 1108 stars in NGC 188 \citep{Geller09}. Assuming the period, mass ratio, and eccentricity distribution from the field solar-type stars we are able to reproduce the velocity dispersion of $\sim 0.5$\,km\,s$^{-1}$ for a single epoch of data, consistent with the value \citet{Geller11} derived from many epochs of radial velocity data, even though about half of the observed stars are fore- and background stars.

For an assumed binary period, mass ratio, and eccentricity distribution this maximum likelihood procedure can be used to characterize the intrinsic velocity distribution for any set of observed radial velocities, where the observed distribution has been significantly broadened by binary orbital motions. This will happen in almost any clusters where the velocity dispersion is lower than several km$\,$s$^{-1}$ such as open clusters, faint Milky Way companion dwarf galaxies, young massive clusters, local star-forming regions, and some low-mass globular clusters. Determining the intrinsic velocity distribution will allow us to constrain the dynamical state of these clusters from a single epoch of radial velocity data, which gives a useful test on our understanding of the current state and evolution of these type of clusters.

\bibliographystyle{astroads}
\bibliography{global2}

\end{document}